%% file: Paper.tex
\documentclass[reprint, superscriptaddress, showpacs, prd]{revtex4-2}
\usepackage{graphicx}
\usepackage{here}
\usepackage{hyperref}
\usepackage[english]{babel}
\usepackage{inputenc}
\usepackage[compact]{titlesec}
\usepackage{url}

\begin{document}
\thispagestyle{empty}

\onecolumngrid

\begin{center}
  \begin{boldmath}
\large{\bf Search for New Hadronic Decays of $\boldmath{h_c}$ and Observation of $\boldmath{h_c\rightarrow K^{+}K^{-}\pi^{+}\pi^{-}\pi^{0}}$}
 \end{boldmath}
\end{center}

\input{Authors.tex}

\begin{center}
\small{(Dated: \today)}
\end{center}

Ten hadronic final states of the $h_c$ decays are investigated via the process $\psi(3686)\rightarrow \pi^0 h_c$, using a data sample of $(448.1 \pm 2.9) \times 10^6$ $\psi(3686)$ events collected with the \mbox{BESIII} detector.
The decay channel $h_c\rightarrow K^{+}K^{-}\pi^{+}\pi^{-}\pi^{0}$ is observed for the first time with a significance of $6.0 \sigma$. The corresponding branching fraction is determined to be $\mathcal{B}(h_c\rightarrow  K^{+}K^{-}\pi^{+}\pi^{-}\pi^{0}) =(3.3 \pm 0.6 \pm 0.6)\times 10^{-3}$ (the first uncertainty is statistical and the second systematical).
Evidence for the decays $h_c\rightarrow \pi^{+} \pi^{-} \pi^{0} \eta$ and $h_c\rightarrow K^{0}_{S}K^{\pm}\pi^{\mp}\pi^{+}\pi^{-}$ is found with a significance of $3.6 \sigma$ and $3.8 \sigma$, respectively. The corresponding branching fractions (and upper limits) are obtained  to be $\mathcal{B}(h_c\rightarrow \pi^{+} \pi^{-} \pi^{0} \eta ) =(7.2 \pm 1.8 \pm 1.3)\times 10^{-3}$ $(< 1.8 \times 10^{-2})$ and $\mathcal{B}(h_c\rightarrow K^{0}_{S}K^{\pm}\pi^{\mp}\pi^{+}\pi^{-}) =(2.8 \pm 0.9 \pm 0.5)\times 10^{-3}$ $(<4.7\times 10^{-3})$. Upper limits on the branching fractions for the final states $h_c\rightarrow K^{+}K^{-}\pi^{0}$, $K^{+}K^{-}\eta$, $K^{+}K^{-}\pi^{+}\pi^{-}\eta$, $2(K^{+}K^{-})\pi^{0}$, $K^{+}K^{-}\pi^{0}\eta$, $K^{0}_{S}K^{\pm}\pi^{\mp}$, and $p\bar{p}\pi^{0}\pi^{0}$ are determined at a confidence level of 90\%.

\maketitle

\twocolumngrid


\section{Introduction}
Although the charmonium spectrum below the open-charm threshold seems to be well understood, it still generates unanswered questions. This is because the charmonium states are located in the transition region of perturbative and non-perturbative quantum chromodynamics (QCD) and theoretical predictions suffer therefore from large uncertainties \cite{Renard, Bodwin, Novikov, Kuang}. The study of charmonium states and their decays is therefore crucial for gaining a deeper understanding of the intermediate-energy regime of QCD, while QCD has been tested successfully at high energies \cite{dissertori2009quantum}.
For the $h_c$, $\chi_{cJ}$ and $\eta_c(2S)$ states, most of the decay channels are still unknown.
After the discovery of the spin-singlet charmonium state $h_c(^1P_1)$ in 2005 \cite{PhysRevD.72.092004,PhysRevLett.95.102003}, there were only few measurements of its decays.
Contrary to the fact that $h_c\rightarrow \gamma \eta_c$ is the prominent decay channel in every calculation, the predictions of the decay process $h_c \rightarrow light\, hadrons$ range from $14-48\%$ \cite{Renard, Bodwin, Novikov, Kuang} depending on the theoretical model.
Therefore experimental measurements are needed to test and improve the theoretical models.\par
Experimental challenges arise from the limited statistics since these non-vector states cannot be produced directly in $e^+ e^-$ annihilation. The best-measured decay mode is the radiative transition $h_c \rightarrow \gamma \eta_c$, occurring in 51\% of all decays \cite{PhysRevD.72.032001,PhysRevLett.101.182003,PhysRevLett.104.132002}, while the sum of all other known branching fractions is less than 3\% \cite{pdg}. Among these measurements, the multi-pionic decay $h_c\rightarrow 2(\pi^{+}\pi^{-})\pi^{0}$ has been confirmed recently by the \mbox{BESIII} collaboration \cite{hctohadrons} after the first evidence was reported by CLEO-c \cite{PhysRevD.80.051106}. Furthermore, \mbox{BESIII} observed the decay mode $h_c\rightarrow p\overline{p}\pi^{+}\pi^{-}$ and reported evidence for the decay $h_c\rightarrow \pi^{+}\pi^{-}\pi^{0}$.
Since the previous analyses mainly studied multi-pionic final states, this analysis focuses on hadronic final states containing kaons as they could lead to intermediate resonances such as e.g. $\phi$ and exited kaon states. After radiative decays of the $h_c$ to $\eta^{(')}$ have been observed, the study of decays involving light vector-states different from the photon will be an extension of these observations. Finally, the observation of $h_c\rightarrow p\overline{p}\pi^{+}\pi^{-}$ motivated us to study the decay with neutral pions, as it could give additional hints on baryonic intermediate states.\par
From these considerations, the following ten final states are chosen to search for undiscovered decay channels of the $h_c$:
 (i) $h_c\rightarrow K^{+}K^{-}\pi^{+}\pi^{-}\pi^{0}$, (ii) $h_c\rightarrow \pi^{+}\pi^{-}\pi^{0} \eta$, (iii) $h_c\rightarrow K^{0}_{S}K^{\pm}\pi^{\mp}\pi^{+}\pi^{-}$, (iv) $h_c\rightarrow K^{+}K^{-}\pi^{0}$, (v) $h_c\rightarrow K^{+}K^{-}\eta$, (vi) $h_c\rightarrow K^{+}K^{-}\pi^{+}\pi^{-}\eta$, (vii) $h_c\rightarrow 2(K^{+}K^{-})\pi^{0}$, (viii) $h_c\rightarrow K^{+}K^{-}\pi^{0}\eta$, (ix) $h_c\rightarrow K^{0}_{S}K^{\pm}\pi^{\mp}$, and (x) $h_c\rightarrow p\bar{p}\pi^{0}\pi^{0}$. These are referenced in this manuscript by roman numbers (i, ii, ..., x).
In this analysis the $h_c$ meson is produced via $\psi(3686)\rightarrow \pi^0 h_c$ using a data sample of $(448.1 \pm 2.9) \times 10^6$ $\psi(3686)$ events \cite{psip_events} collected with the \mbox{BESIII} detector.

\section{BESIII Detector and Monte Carlo Simulation}

The \mbox{BESIII} detector is a magnetic spectrometer~\cite{BES_Detector} located at the Beijing Electron Positron Collider (BEPCII)~\cite{Yu:IPAC2016-TUYA01}. The cylindrical core of the \mbox{BESIII} detector consists of a helium-based multilayer drift chamber (MDC), a plastic scintillator time-of-flight system (TOF), and a CsI(Tl) electromagnetic calorimeter (EMC), which are all enclosed in a superconducting solenoidal magnet providing a 1.0~T magnetic field. The solenoid is supported by an octagonal flux-return yoke with resistive plate counter muon identifier modules interleaved with steel. The acceptance of charged particles and photons is 93\% over the $4\pi$ solid angle. The charged-particle momentum resolution at $1~ \mathrm{GeV}/c$ is $0.5\%$, and the $dE/dx$ resolution is $6\%$ for the electrons from Bhabha scattering. The EMC measures photon energies with a resolution of $2.5\%$ ($5\%$) at $1$~GeV in the barrel (end cap) region. The time resolution of the TOF barrel part is 68~ps, while that of the end cap part is 110~ps.\par

{\sc geant4}-based \cite{evtgen1,evtgen2} Monte Carlo (MC) simulations are used to study the detector response and to estimate background contributions. Inclusive MC samples are produced to estimate the contributions from possible background channels. The production of the initial $\psi(3686)$ resonance in $e^+e^-$ annihilation is simulated using the MC event generator {\sc kkmc} \cite{kkmc1,kkmc2}. Its known decay modes are modeled with {\sc evtgen} \cite{evtgen1,evtgen2} using the world average branching fraction values \cite{pdg16}, while the remaining unknown decays are generated using {\sc lundcharm} \cite{lundcharm}.
{\sc geant4} is used to simulate the particle propagation through the detector system.
The simulation includes the beam-energy spread and initial-state radiation (ISR) in the $e^+e^-$ annihilations.
In addition, exclusive MC samples containing one million events are generated using the phase-space model (PHSP) for each signal mode to optimize the selection criteria and to study the efficiency.

\begin{table}[htb]
\centering
\caption{Applied requirements on the $\chi^{2}_{(4+N)\mathrm{C}}$ and mass windows used as vetoes in each exclusive mode. The lower case $m$ denotes the nominal particle mass \citep{pdg}.}
\label{tab:summary_vetoes}
\begin{ruledtabular}
\begin{tabular}{ccr@{\hspace{-1.5em}}l}

Mode & $\chi^{2}_{(4+N)\mathrm{C}}$ limit & \multicolumn{2}{c}{Mass Windows $[\mathrm{MeV/}c^{2}]$}\\
\hline
(i) & $<60$ & $| M (\pi^{+} \pi^{-})_{rec} -m_{J/\psi} |$ & $> 25$\\
& & $| M (\pi^{0} \pi^{0})_{rec} -m_{J/\psi} |$ & $> 25$ \\
& &  $| M (\pi^{+} \pi^{-} \pi^{0}_ {1}) - m_{\omega} |$ & $> 20$\\
& & $| M (\pi^{+} \pi^{-} \pi^{0}_ {1}) - m_{\eta} |$ & $> 16$ \\
& & $820 < M(K^{\pm}\pi^{0}_ {1})$ & $< 920$ \\
\hline
(ii) & $<100$ & $| M (\pi^{+} \pi^{-})_{rec} -m_{J/\psi} |$ & $> 30$\\
& & $| M(\eta)_{rec} -m_{J/\psi} |$ & $> 30$\\
& & $| M (\pi^{+} \pi^{-} \pi^{0}_ {1}) - m_{\omega} |$ & $> 20$\\
& & $| M (\pi^{+} \pi^{-} \pi^{0}_ {1}) - m_{\eta} |$ & $> 16$ \\
\hline
(iii) & $<40$ & $| M (\pi^{+} \pi^{-})_{rec} - m_{J/\psi} |$ & $> 30$\\
& & $| M (\pi^{+} \pi^{-} \pi^{0}_ {1}) - m_{\omega} |$ & $>20$\\
& & $| M (\pi^{+} \pi^{-} \pi^{0}_ {1}) - m_{\eta} |$ & $>20$ \\
& & $| M (K_{S}^{0}\pi^{0}_{1}) -  m_{K^{*}} |$ & $>50$\\
& & $| M (K^{\pm}\pi^{0}_{1}) -  m_{K^{*}} |$ & $>50$\\
\hline
(iv) & $<100$ & $| M (\pi^{0} \pi^{0})_{rec} - m_{J/\psi} |$ & $>30$ \\
& & $| M (K^{\pm}\pi^{0}_{1}) -  m_{K^{*}} |$ &  $> 50$\\
\hline
(v) & $<100$ &  & \\
\hline
(vi) & $<60$ & $| M (\pi^{+} \pi^{-})_{rec} - m_{J/\psi} |$ & $>30$\\
& & $| M(\eta)_{rec} -m_{J/\psi} |$  & $>30$\\
& & $| M (K^{\pm}\pi^{0}_{1}) -  m_{K^{*}} |$ & $>50$\\
\hline
(vii) & $<100$ & $| M (\pi^{+} \pi^{-})_{rec} -m_{J/\psi} |$ & $>30$\\
& & $| M (K^{\pm}\pi^{0}_{1}) -  m_{K^{*}} |$ & $>20$\\
\hline
(viii) & $<100$ & $| M (\pi^{+} \pi^{-})_{rec} - m_{J/\psi} |$ & $>30$\\
& & $| M(\eta)_{rec} - m_{J/\psi} |$ & $>30$\\
\hline
(ix) & $<100$ & $| M (\pi^{+} \pi^{-})_{rec} - m_{J/\psi} |$ & $>30$\\
& & $| M (K_{S}^{0}\pi^{0}_{1}) -  m_{K^{*}} |$ & $>50$\\
& & $| M (K^{\pm}\pi^{0}_{1}) -  m_{K^{*}} |$ & $>50$\\
\hline
(x) & $<50$ & $| M (\pi^{0} \pi^{0})_{rec} - m_{J/\psi} |$ & $>30$ \\
& & $|M(p \pi^{0}_{1}) - m_{\Delta(1232)^{+}}|$ & $>10$ \\
& & $|M(p \pi^{0}_{1}) - m_{\Sigma^{+}}|$ & $>30$ \\
& & $|M (\pi^{0}\pi^{0}\pi^{0}) - m_{\eta} |$ & $>25$ \\

\end{tabular}
\end{ruledtabular}
\end{table}

\begin{figure*}[htb]
\includegraphics[scale=0.3]{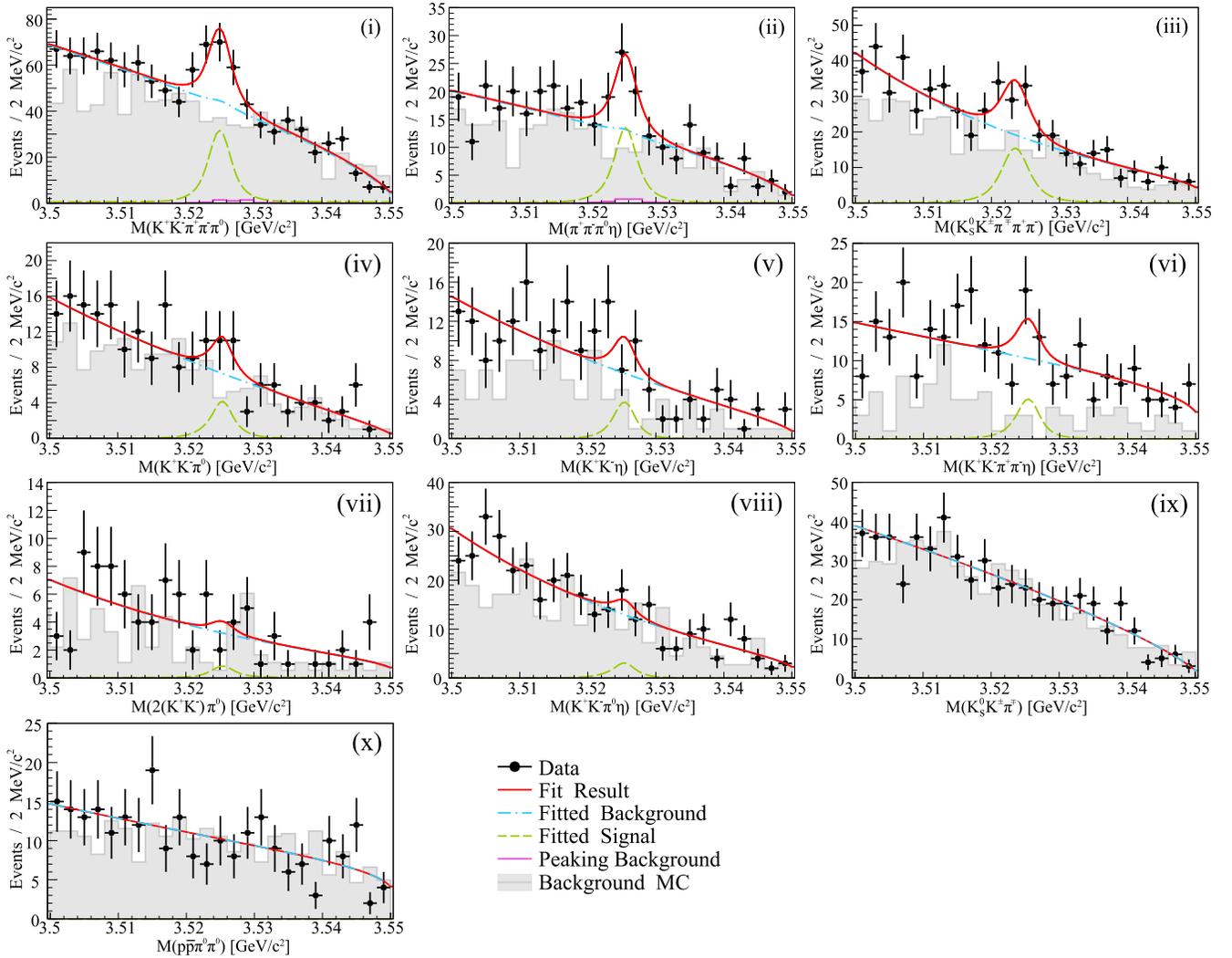}
\caption{Fits to the invariant mass distributions for the $h_c$ decay modes (i)-(x). Data are shown as black points, the total fit result is shown in red, the background contribution is denoted by the blue dashed-dotted line (including peaking background contributions for channel (i) and (ii) as shown in magenta), the signal contribution is illustrated by the green dashed line. The background level obtained from inclusive MC is shown by the gray shaded histogram.}
\label{fig:fit_results}
\end{figure*}

\section{Data Analysis}
Each charged track reconstructed in the MDC is required to originate from a region of 10 cm of the interaction point along the beam direction and 1 cm in the plane perpendicular to the beam. The polar angle $\theta$ of the tracks must be within the fiducial volume of the MDC $\mathrm{|cos \theta| < 0.93}$. Tracks used in reconstructing $K_{S}^{0}$ mesons are exempted from these requirements and $\mathrm{|cos \theta| < 0.93}$ is required only for the daughter pions.
The TOF and $dE/dx$ measurements for each charged track are combined to compute particle identification (PID) confidence levels for pion, kaon and proton hypotheses. The track is assigned to the particle type with the highest confidence level which has to be larger than 0.001.\par
Photon candidates are reconstructed from electromagnetic showers produced in the crystals of the EMC.
A shower is treated as a photon candidate if the deposited energy is larger than $25\, \mathrm{MeV}$ in the barrel region $\mathrm{(|cos\theta| < 0.8)}$ or 50 $\mathrm{MeV}$ in the end cap region $\mathrm{(0.86 < | cos \theta| < 0.92)}$. The timing of the shower is required to be within 700 ns from the reconstructed event start time to suppress noise and energy deposits unrelated to the event. To remove
Bremsstrahlung photons, the angle between the photon and the extrapolated impact point in the EMC of the nearest charged track must be larger than $10^{\circ}$ for charged pions and kaons and $20^{\circ}$ for protons, respectively.\par
Following the application of a vertex fit that constraints all charged tracks to arise from a common interaction point, a kinematic fit, constraining the total energy and momentum to the initial four momentum, is performed to further improve the momentum resolution and to suppress background. Final states containing a $K_{S}^{0}$ undergo a secondary vertex fit, which ensures the daughter pions being produced in a common vertex. A $K_{S}^{0}$ candidate is accepted if $487 <M(\pi^{+}\pi^{-})<511 \,\mathrm{MeV/}c^2$ and $\lambda/\Delta\lambda>2$, where $\Delta \lambda$ is the uncertainty on the decay length $\lambda$ obtained from the secondary vertex fit.\par
A pair of photons is treated as a $\pi^{0}$ or $\eta$ candidate if it satisfies $|M(\gamma\gamma)-M_{\pi^{0}}|<30\,\mathrm{MeV/}c^{2}$ or $|M(\gamma\gamma)-M_{\eta}|<50\,\mathrm{MeV/}c^{2}$, which corresponds to an interval of about $\pm 3$ times the mass resolution. In all final states containing $\pi^{0}$ or $\eta$ mesons, the kinematic fit also constraints $\gamma \gamma$ pairs to have the expected nominal masses. The combination with the best $\chi^{2}$ value is kept for the further analysis in case there are more than one $\gamma\gamma$ combinations.\par
To suppress contamination from decays with different numbers of photons, such as the dominant decay $\psi(3686)\rightarrow\gamma \chi_{c2}$, where the $\chi_{c2}$ decays to the same final states as the $h_c$, the following procedure is applied. The $\chi^{2}_{4\mathrm{C}, n\gamma}$ value is obtained from a four-constraint fit including the expected number of photons $n$ for a given signal hypothesis with respect to the initial four momentum. The value $\chi^{2}_{4\mathrm{C}, (n-1)\gamma}$ is determined from an additional 4C fit with one missing photon compared to the desired signal process. An event is rejected if $\chi^{2}_{4\mathrm{C}, n\gamma} > \chi^{2}_{4\mathrm{C}, (n-1) \gamma}$. To further suppress background, the $\chi^{2}_{(4+N)\mathrm{C}}$ value of the total kinematic fit, including additional mass constraints for $\pi^{0}$, $\eta$ and $K_{S}^{0}$ candidates (denoted by N), is limited depending on the final state (see Table~\ref{tab:summary_vetoes}). Additional vetoes in the $\pi^{0} \pi^{0}, \, \pi^{+} \pi^{-}$ and $\eta$ recoil masses, as listed in Table~\ref{tab:summary_vetoes}, are applied to suppress background from $\psi(3686)\rightarrow (\pi^{0} \pi^{0}, \, \pi^{+} \pi^{-}, \, \eta ) \, J/\psi$.
Since the $\pi^{0}$ from the decay $\psi(3686)\rightarrow \pi^{0} h_c$ (denoted by $\pi^{0}_{1}$ and identified among all $\pi^{0}$ candidates by its energy being closest to the expected) should not create any structure together with other final state particles. Therefore, additional vetoes are applied to suppress background from $\omega\rightarrow \pi^{+}\pi^{-}\pi^{0}$, $\eta\rightarrow \pi^{+}\pi^{-}\pi^{0}$, $\eta\rightarrow \pi^{0}\pi^{0}\pi^{0}$, $K^{*\pm}\rightarrow \pi^{0} K^{\pm}$, $\Delta(1232)^{+}\rightarrow p\pi^{0}$ and $\Sigma^{+}\rightarrow p\pi^{0}$  as given in Table~\ref{tab:summary_vetoes}. The mass windows for these vetoes and for the $\chi^2$ criterion are optimized simultaneously for each channel using the figure of merit of $S/\sqrt{ S+B}$ and are listed in Table~\ref{tab:summary_vetoes}. Here, $S$ denotes the number of signal events, obtained from signal MC, which is scaled to the branching fraction as determined in this analysis. Therefore, the unoptimized selection criteria were used in a first iteration to obtain a preliminary branching fraction (or upper limit). This preliminary result is then fed into the next iteration of optimization until the procedure converges within the uncertainties. The number of background events $B$ is obtained from the $\psi(3686)$ inclusive MC and scaled to the expected number of events. Figure~\ref{fig:fit_results} shows the obtained invariant mass distributions of the different decay modes. After applying all selection criteria, the remaining background originates mostly from the non-resonant production of the same final state particles as the signal and thus cannot be suppressed further.

\section{Determination of Branching Fractions}

\begin{table*}[htb]
\centering
\caption{Overview of the branching fractions and upper limits obtained in this analysis for decay processes of the $h_c$ meson. The first uncertainty shown is the statistical and the second the systematical uncertainty of the measurement method which includes the uncertainty that arises due to the use of external branching fractions.}

\label{tab:summary_bf}
\begin{ruledtabular}
\begin{tabular}{c c c c c c}
Mode & $X$ & $N_{h_c}$ & $\varepsilon\, (\%)$ & $\mathcal{B}(\psi(3686) \rightarrow \pi^{0} h_c) \times \mathcal{B}(h_c\rightarrow X)$ & $\mathcal{B}(h_c\rightarrow X)$ \\
\hline
(i)     & $K^{+}K^{-}\pi^{+}\pi^{-}\pi^{0}$ & $80 \pm 15$ & $6.5$ & $(2.8 \pm 0.5 \pm 0.3)\times 10^{-6}$ & $(3.3 \pm 0.6 \pm 0.6)\times 10^{-3}$ \\

(ii)    & $\pi^{+}\pi^{-}\pi^{0} \eta$ & $35 \pm 9$ & $3.3$ &$(6.2 \pm 1.6 \pm 0.7)\times 10^{-6}$ & $(7.2 \pm 1.8 \pm 1.3)\times 10^{-3}$\\
        & & $<50.0$ & &  $\mathrm{< 1.5 \times 10^{-5}}$ & $\mathrm{< 1.8 \times 10^{-2}}$ \\

(iii)   & $K^{0}_{S}K^{\pm}\pi^{\mp}\pi^{+}\pi^{-}$ & $41 \pm 13$ & $5.5$ & $(2.4 \pm 0.7 \pm 0.3)\times 10^{-6}$ & $(2.8 \pm 0.9 \pm 0.5)\times 10^{-3}$ \\
        & & $<65.3$ & &  $\mathrm{<3.9 \times 10^{-6}}$ & $\mathrm{<4.7 \times 10^{-3}}$ \\

(iv)    & $K^{+}K^{-}\pi^{0}$ & $<20.1$ & $9.8$ & $\mathrm{<4.8 \times 10^{-7}}$ & $\mathrm{<5.8 \times 10^{-4}}$ \\

(v)     & $K^{+}K^{-}\eta$ & $<18.5$ & $14.3$  & $\mathrm{<7.5 \times 10^{-7}}$ & $\mathrm{<9.1 \times 10^{-4}}$\\

(vi)    & $K^{+}K^{-}\pi^{+}\pi^{-}\eta$ & $<24.1$ & $6.9$ & $\mathrm{<2.0 \times 10^{-6}}$ & $\mathrm{<2.5 \times 10^{-3}}$ \\

(vii)   & $2(K^{+}K^{-})\pi^{0}$ & $<11.7$ & $6.7$ & $<2.1 \times 10^{-7}$ & $\mathrm{<2.5 \times 10^{-4}}$  \\

(viii)  & $K^{+}K^{-}\pi^{0}\eta$ & $<20.2$ & $6.3$ & $\mathrm{<1.8 \times 10^{-6}}$ & $\mathrm{<2.2 \times 10^{-3}}$  \\

(ix)    & $K^{0}_{S}K^{\pm}\pi^{\mp}$ & $<17.4$ & $14.4$ & $\mathrm{<4.8 \times 10^{-7}}$ & $\mathrm{<5.7 \times 10^{-4}}$ \\

(x)     & $p\bar{p}\pi^{0}\pi^{0}$ & $<11.8$ & $8.7$  & $\mathrm{<4.4 \times 10^{-7}}$ & $\mathrm{<5.2 \times 10^{-4}}$ \\

\end{tabular}
\end{ruledtabular}
\end{table*}

To determine the number of signal events $N_{h_c}$, an unbinned maximum likelihood fit to the invariant mass spectra of the particles to reconstruct the $h_c$ is performed as shown in Fig.~\ref{fig:fit_results}. In each fit, the signal contribution is described by a Breit-Wigner function convoluted with a detector resolution function as given in Ref. \cite{Das2016}. Here, the mass and width of $h_c$ in the Breit-Wigner function are fixed to their world average values \cite{pdg}, and the parameters in the resolution function are determined with the signal MC simulation. The background shape is described by an ARGUS function \cite{ALBRECHT1994217}, where the threshold parameter of the ARGUS function is fixed to the kinematical threshold of $3551 \, \mathrm{MeV}/c^2$. In case of the modes $h_c\rightarrow K^{+}K^{-}\pi^{+}\pi^{-}\pi^{0}$ and $h_c\rightarrow \pi^{+} \pi^{-} \pi^{0} \eta$, additional peaking background from the processes $h_c\rightarrow \gamma \eta_c,\, \eta_c\rightarrow K^{+}K^{-}\pi^{+}\pi^{-}$ and $\eta_c\rightarrow \pi^{+}\pi^{-} \eta$ is included and scaled to the expected number of events based on the world average values.
The resulting branching fractions are determined by:

\begin{equation}
\mathcal{B}(h_c \rightarrow X)\cdot \mathcal{B}(\psi(3686)\rightarrow \pi^{0} h_c)=\frac{N_{h_c}}{N_{\psi(3686)}  \cdot \prod_i\mathcal{B}_i \cdot \varepsilon}.
\end{equation}

\noindent Here $\mathcal{B}(h_c \rightarrow X)$ denotes the branching fraction of the $h_c$ meson decaying to final state $X$, the branching fraction of $\psi(3686)\rightarrow \pi^{0} h_c$ is given by $\mathcal{B}(\psi(3686)\rightarrow \pi^{0} h_c) = (8.6 \pm 1.3)\times 10^{-4}$ \cite{pdg}. The number of $\psi(3686)$ events is given by $N_{\psi(3686)}=(448.1 \pm 2.9) \times 10^6$ \cite{psip_events}. And $\prod_i\mathcal{B}_i$ is the product of branching fractions of the decaying particles like $\mathcal{B}(\pi^{0}\rightarrow \gamma \gamma)$, $\mathcal{B}(\eta\rightarrow \gamma \gamma)$ and $\mathcal{B}(K_{S}^{0}\rightarrow \pi^{+}\pi^{-})$ taken from \cite{pdg}. The efficiency $\varepsilon$ is obtained from signal MC simulations and $N_{h_c}$ is the number of signal events obtained by the fit. In order to determine $\mathcal{B}(h_c \rightarrow X)$, the product branching fraction is divided by $\mathcal{B}(\psi(3686)\rightarrow \pi^{0} h_c)$. Both values, the product branching fraction and the $h_c$ decay branching fraction are listed in Table~\ref{tab:summary_bf}.

In case no significant signal contribution is observed for the modes (iv)-(x), upper limits on the branching fractions are determined by the Bayesian approach \cite{Demortier:2002ic}. To obtain the likelihood distribution, the signal yield is scanned using the fit function described earlier. Systematic uncertainties are considered by smearing the obtained likelihood curve with a Gaussian function with the width of the systematic uncertainty of the respective decay mode. The upper limit at a confidence level of 90\% is finally obtained by:
\begin{equation}
0.9 = \frac{\int_{0}^{N_{h_c}^{up}} dN \mathcal{L}(N) }{\int_{0}^{\infty} dN \mathcal{L}(N)}.
\end{equation}
The upper limit on the number of observed events $N_{h_c}^{up}$ is determined by integrating the smeared likelihood function $\mathcal{L}(N)$ up to the value $N_{h_c}^{up}$, which corresponds to 90\% of the integral. The results are listed in Table~\ref{tab:summary_bf}.\par

 Among the ten final states,
 the decay $h_c\rightarrow K^{+}K^{-}\pi^{+}\pi^{-}\pi^{0}$ is observed with a statistical significance of $6 \sigma$ and evidences for the decays $h_c\rightarrow \pi^{+} \pi^{-} \pi^{0} \eta$ and $h_c\rightarrow K^{0}_{S}K^{\pm}\pi^{\mp}\pi^{+}\pi^{-}$ are found with statistical significances of $3.6 \sigma$ and $3.8 \sigma$, respectively. The combined significance of the modes (iv)-(x) is determined to be $3.5 \sigma$. The statistical significance is determined by the likelihood ratio between a fit with and without signal component by taking the change in the number of fit parameters into account.

\begin{figure}
\includegraphics[width=1. \columnwidth]{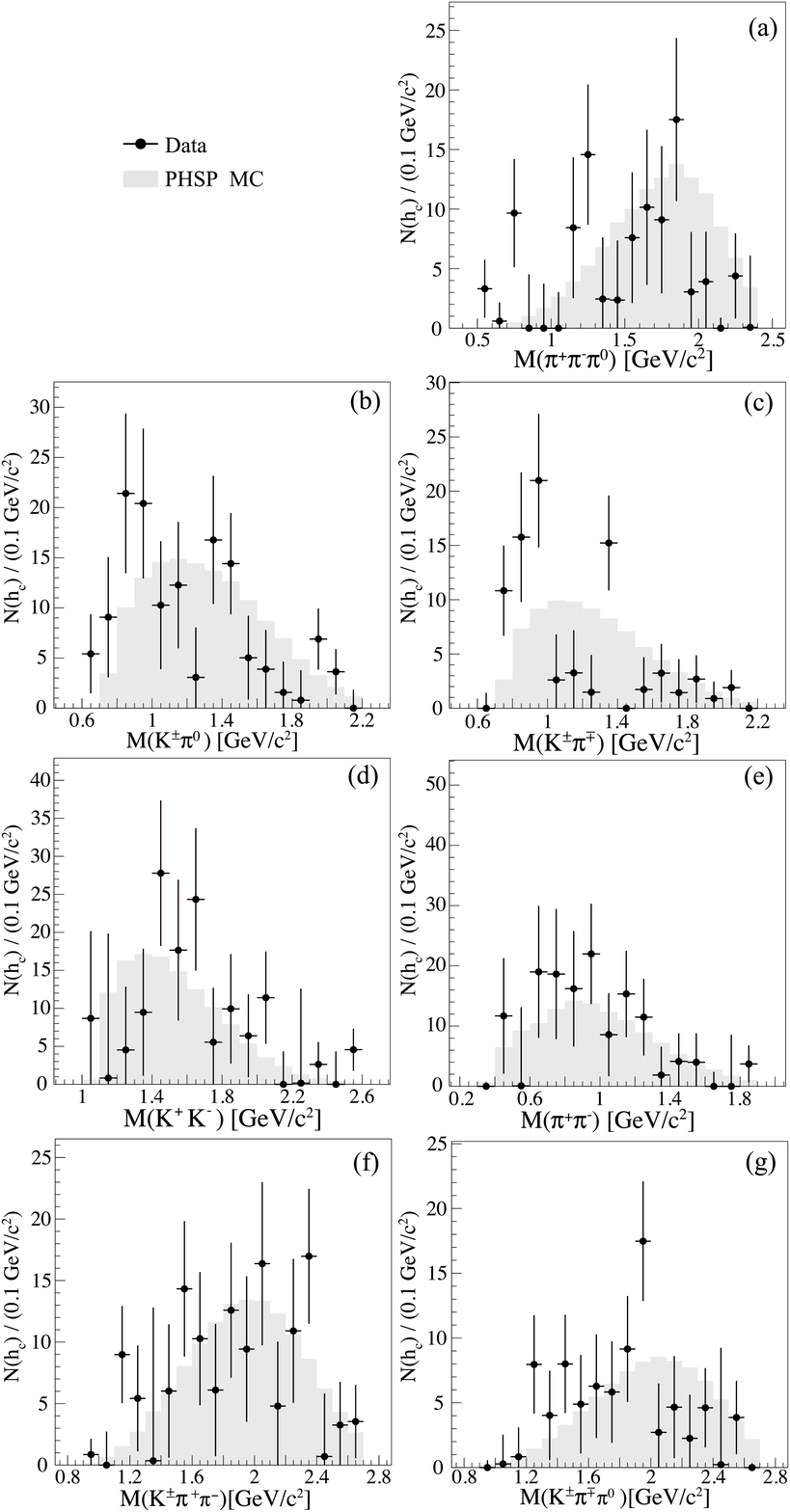}
\caption{Signal yield obtained from an unbinned maximum-likelihood fit to the invariant $K^{+}K^{-}\pi^{+}\pi^{-}\pi^{0}$ mass in slices of the invariant  $\pi^{+}\pi^{-}\pi^{0}$ (a), $K^{\pm}\pi^{0}$ (b), $K^{\pm}\pi^{\mp}$ (c), $K^{+}K^{-}$ (d), $\pi^{+}\pi^{-}$ (e), $K^{\pm}\pi^{+}\pi^{-}$ (f) and $K^{\pm}\pi^{\mp}\pi^{0}$ (g) mass. Black dots denote the signal yield determined from data. The grey shaded histogram shows the PHSP distribution obtained from MC which is scaled to the integral of signal yield.}
\label{fig:subsystem}
\end{figure}

For the final state $K^{+}K^{-}\pi^{+}\pi^{-}\pi^{0}$ in the $h_c$ decay, a search for intermediate resonances is performed to obtain information about underlying sub-processes. Despite the large background contamination, the signal content is determined by an unbinned maximum likelihood fit to the invariant $K^{+}K^{-}\pi^{+}\pi^{-}\pi^{0}$ mass in slices of masses of possible sub-systems. The resulting distributions are shown in the Fig.~\ref{fig:subsystem}.

No firm conclusions about contributions of intermediate resonances can be drawn based only on the extracted projections with the present statistics. The $K\pi$ distribution shows a possible structure in the $K^{*}(892)$ region, which may signal the production of this resonance. In the invariant $K^{\pm}\pi^{\mp}\pi^{0}$ mass distribution there may be a hint in the mass region of $1.9 - 2.0 \, \mathrm{GeV/}c^{2}$ for the production of an excited kaon, such as the $K_{2}^{*}(1980)$ or $K_{2}(1820)$. Conceivable sub-processes would then be $h_c \rightarrow \left( K^{*}(892)/K_{0,2}^{*}(1430)\right) \left( K_2(1820)/K_2^{*}(1980)\right)$. As shown in this analysis, evidence for the decay $h_c\rightarrow K_{S}^{0}K^{\pm}\pi^{\mp}\pi^{+}\pi^{-}$ has been found.

\section{Systematic Uncertainties}

The sources of systematic uncertainties for the branching fractions include tracking, PID, selection, uncertainties caused by the signal fitting procedure and efficiency determination, and they are explained in the following. All the systematic uncertainties are summarized in Table~\ref{tab:summary_sys}. The overall systematic uncertainty for the product branching fraction $\mathcal{B}(\psi(3686) \rightarrow \pi^{0} h_c) \times \mathcal{B}(h_c\rightarrow X)$ is obtained by summing all individual components (from column two to six of Table~\ref{tab:summary_sys}) in quadrature. An additional systematic uncertainty of $\Delta_{ext}=$15.1\% is added in quadrature due to the branching fraction of $\psi(3686)\rightarrow \pi^{0} h_c$ used in the calculation of branching fractions of $\mathcal{B}(h_c\rightarrow X)$.\par
The uncertainties of the tracking efficiency are estimated using $J/\psi\rightarrow p\bar{p}\pi^{+}\pi^{-}$ and $e^{+}e^{-}\rightarrow \pi^{+}\pi^{-}K^{+}K^{-}$ control samples \cite{p_tracking, k_pi_tracking}. The resulting uncertainties are determined to be 1\% for each participating charged pion, kaon, proton, and antiproton, of a particular final state. The uncertainties due to PID are studied with control samples $e^{+}e^{-}\rightarrow \pi^{+}\pi^{-}K^{+}K^{-}$ and $e^{+}e^{-}\rightarrow p \bar{p}\pi^{0}$ and are estimated to be 1\% for each participating pion, kaon, proton and anitproton \cite{k_pi_tracking,pion_PID}. The uncertainty due to photon reconstruction is estimated to be 1\% per photon based on studies of the reference channel $J/\psi \rightarrow \rho^0\pi^{0}$ \cite{photon_effi}. The uncertainties due to $\pi^{0},\, \eta$ and $K_{S}^{0}$ reconstruction are studied by using the reference processes $J/\psi\rightarrow \pi^{+}\pi^{-}\pi^{0}$, $J/\psi \rightarrow \eta p \bar{p} $ and $J/\psi \rightarrow K^{*\pm}(892) K^{\mp}, \, K^{*\pm}(892)\rightarrow K_{S}^{0} \pi^{\pm}$ and are estimated to be 1\% per $\pi^{0}$ and $\eta$ \cite{photon_effi} and 1.2\% per $K_{S}^{0}$ \cite{ks_reco}. For example in case of final state (i), two pions, two kaons and four photons reconstructed as $\pi^{0}$ are involved. This gives the following contributions to the systematic uncertainty: 4\% (1\% per track) due to PID, 4\% (1\% per track) due to track reconstruction, 4\% due to photon reconstruction (1\% per photon) and additional 2\% for $\pi^{0}$ reconstruction (1\% for each $\pi^{0}$).
The total uncertainty due to PID and event reconstruction is given for each final state in the second column of Table~\ref{tab:summary_sys}. \par
The systematic uncertainty due to the selection criteria is determined by varying the nominal selection criteria. For each mass window requirement, the nominal value of the criterion is varied by $\pm 10 \,\mathrm{MeV/}c^2$ in increments of $\pm 0.5 \, \mathrm{MeV/}c^2$. The maximum deviation from the nominal branching fraction is quoted as a systematic uncertainty as given in column three of Table~\ref{tab:summary_sys}.\par
The uncertainties associated with the kinematic fit are determined by comparing the efficiencies with and without the helix parameter correction. For charged particles, differences in the $\chi^{2}$ distributions of the kinematic fit between data and MC have been studied by using a control sample $J/\psi\rightarrow \phi f_{0}(980)$, $\phi\rightarrow K^+K^-$, $f_{0}(980)\rightarrow \pi^+\pi^-$, which ensures high statistics and purity \cite{helix_cor}. The helix parameters of the corresponding tracks are corrected accordingly \cite{helix_cor}. The difference between the result determined with and without this correction applied are assigned as systematic uncertainty of the kinematic fit.\par
The systematic uncertainty due to the physics model and the efficiency used to simulate signal MC arises from the limited knowledge of intermediate states in the $h_c$ decay. Therefore have MC samples been generated including additional intermediate states and the results are compared with those of the nominal phase space sample.
The systematic uncertainty of the fit results from the choice of background parametrization is determined by using a second order Chebychev polynomial. In case of a peaking background, the uncertainty of the branching fraction (e.g. $\mathcal{B}(\eta_c\rightarrow\pi^{+}\pi^{-}K^{+}K^{-}) = (6.9 \pm 1.1) \times 10^{-3}$) \cite{pdg} has been used to determine the uncertainty of the peaking background. This uncertainty contributes 1.2\% in case of $h_c\rightarrow K^{+}K^{-}\pi^{+}\pi^{-}\pi^{0}$ and 2\% in $h_c\rightarrow \pi^{+} \pi^{-} \pi^{0} \eta$, respectively. Another contribution to the systematic uncertainty of the fit is caused by the limited range of the invariant mass in which the fit is applied. Therefore, the fit range is extended from $3.50 - 3.55 \, \mathrm{GeV/}c^{2}$ to $3.40 - 3.65 \, \mathrm{GeV/}c^{2}$, and the difference of the branching fraction result is used as a systematic uncertainty. Further uncertainties arise from the parametrization of the resolution distributions. Instead of using the default parametrization, a Crystal Ball distribution has been used.
The uncertainty arising from the number of $\psi(3686)$ events is 0.7\% \cite{psip_events}.

\begin{table}[htb]
\centering
\caption{Summary of systematic uncertainties. Here $\Delta_{\mathcal{BB}}=\sqrt{\Sigma_i\Delta_{i}^2}$ is the systematic uncertainty for $\mathcal{B}(\psi(3686) \rightarrow \pi^{0} h_c) \times \mathcal{B}(h_c\rightarrow X)$ and $\Delta_{\mathcal{B}}=\sqrt{\Sigma_i\Delta_{i}^2+\Delta_{ext}^2}$ is that for $\mathcal{B}(h_c\rightarrow X)$ . $\Delta_{i}$ represents the individual uncertainties given in columns two to six and $\Delta_{ext}$=15.1\% is an additional uncertainty for $\mathcal{B}(h_c\rightarrow X)$ due to the external uncertainty of $\mathcal{B}(\psi(3686) \rightarrow \pi^{0} h_c)$ \citep{pdg}. All values are given in \%.}
\label{tab:summary_sys}
\begin{ruledtabular}
\begin{tabular}{c c c c c c c c}

Mode & PID, & Selec- & Kin. & Eff. & Fit & $\Delta_{\mathcal{BB}}$ & $\Delta_{\mathcal{B}}$ \\
& Reco. & tion & Fit& & & \\
\hline
  (i)     & 7.2 & 5.0 & 1.5 & 5.1 & 2.7 & 10.6 & 18.4 \\
  (ii)    & 7.0 & 4.4 & 2.1 & 6.3 & 2.8 & 11.0 & 18.7 \\
  (iii)   & 8.9 & 3.9 & 3.9 & 5.3 & 4.0 & 12.4 & 19.5 \\
  (iv)    & 5.3 & 3.7 & 1.9 & 4.1 & 3.8 & 8.8  & 17.5 \\
  (v)     & 5.1 & 2.0 & 1.9 & 3.7 & 3.4 & 7.7  & 16.9 \\
  (vi)    & 7.1 & 4.1 & 1.6 & 3.4 & 4.7 & 10.2 & 18.2 \\
  (vii)   & 7.2 & 4.7 & 2.3 & 5.2 & 3.7 & 10.6 & 18.5 \\
  (viii)  & 7.0 & 3.7 & 2.0 & 4.5 & 4.0 & 10.1 & 18.2 \\
  (ix)    & 6.2 & 3.3 & 2.1 & 4.0 & 3.4 & 9.0  & 17.6 \\
  (x)     & 7.3 & 6.0 & 2.7 & 4.9 & 3.3 & 11.5 & 19.0 \\

\end{tabular}
\end{ruledtabular}
\end{table}

\section{Summary}
In this analysis, ten final states of the $h_c$ decays have been searched for using a data sample of $(448.1 \pm 2.9)\times 10^6$  $\psi(3686)$ events collected at \mbox{BESIII}. The decay $h_c\rightarrow K^{+}K^{-}\pi^{+}\pi^{-}\pi^{0}$ is observed for the first time.
Furthermore, evidence for the decays $h_c\rightarrow \pi^{+} \pi^{-} \pi^{0} \eta$ and $h_c\rightarrow K^{0}_{S}K^{\pm}\pi^{\mp}\pi^{+}\pi^{-}$ are found with statistical significances of $3.6 \sigma$ and $3.8 \sigma$, respectively. The combined significance of the modes (iv)-(x) is determined to be $3.5 \sigma$. Upper limits are determined in case there was no signal observed. The measured branching fractions and upper limits at 90\% confidence level are listed in Table~\ref{tab:summary_bf}. \par
Summing up the branching fractions obtained in this analysis, shows that these decays contribute at a level of $\sim 1.3\%$ to all decays and contribute at the same level as the previously observed decays $h_c\rightarrow 2(\pi^{+}\pi^{-})\pi^{0}$ and $h_c\rightarrow p\overline{p}\pi^{+}\pi^{-}$ \cite{hctohadrons}. After the previous observations of multi-pionic decays \cite{hctohadrons}, this is the first observation of the $h_c$ decaying to mesons carrying strangeness. This observation adds another decay mode to the few observed hadronic decays of the $h_c$ and the calculated upper limits further rule out strong contributions of other promising decay channels. These measurements provide input to theoretical models in order to improve their predictions in the future. Finally, it is still unclear if the hadronic decays width of the $h_c$ is of the same order as the radiative decay width predicted in \cite{Bodwin}, or if the radiative decays dominate. Although many final states have been investigated in this analysis, using the largest available data set of resonantly produced $\psi(3686)$ events, future experimental measurements of higher precision together with improved theoretical calculations can further contribute to answering these questions~\cite{Ablikim:2019hff}.

\section{Acknowledgments}
The \mbox{BESIII} collaboration thanks the staff of BEPCII and the IHEP computing center for their strong support. This work is supported in part by National Key Basic Research Program of China under Contract No. 2015CB856700; National Natural Science Foundation of China (NSFC) under Contracts Nos. 11625523, 11635010, 11735014, 11822506, 11835012; the Chinese Academy of Sciences (CAS) Large-Scale Scientific Facility Program; Joint Large-Scale Scientific Facility Funds of the NSFC and CAS under Contracts Nos. U1532257, U1532258, U1732263, U1832207; CAS Key Research Program of Frontier Sciences under Contracts Nos. QYZDJ-SSW-SLH003, QYZDJ-SSW-SLH040; 100 Talents Program of CAS; INPAC and Shanghai Key Laboratory for Particle Physics and Cosmology; ERC under Contract No. 758462; German Research Foundation DFG under Contracts Nos. Collaborative Research Center CRC 1044, FOR 2359; Istituto Nazionale di Fisica Nucleare, Italy; Ministry of Development of Turkey under Contract No. DPT2006K-120470; National Science and Technology fund; STFC (United Kingdom); The Knut and Alice Wallenberg Foundation (Sweden) under Contract No. 2016.0157; The Royal Society, UK under Contracts Nos. DH140054, DH160214; The Swedish Research Council; U. S. Department of Energy under Contracts Nos. DE-FG02-05ER41374, DE-SC-0010118, DE-SC-0012069; University of Groningen (RuG) and the Helmholtzzentrum f\"ur Schwerionenforschung GmbH (GSI), Darmstadt.
\bibliography{refs}%

\end{document}

%% file: Authors.tex
\begin{center}
M.~Ablikim$^{1}$, M.~N.~Achasov$^{10,c}$, P.~Adlarson$^{67}$, S. ~Ahmed$^{15}$, M.~Albrecht$^{4}$, R.~Aliberti$^{28}$, A.~Amoroso$^{66A,66C}$, M.~R.~An$^{32}$, Q.~An$^{63,49}$, X.~H.~Bai$^{57}$
% ...
\\
\vspace{0.2cm}
(BESIII Collaboration)\\
\vspace{0.2cm} {\it
$^{1}$ Institute of High Energy Physics, Beijing 100049, People's Republic of China\\
% ...
$^{4}$ Bochum Ruhr-University, D-44780 Bochum, Germany\\
% ...
$^{10}$ G.I. Budker Institute of Nuclear Physics SB RAS (BINP), Novosibirsk 630090, Russia\\
% ...
$^{15}$ Helmholtz Institute Mainz, Johann-Joachim-Becher-Weg 45, D-55099 Mainz, Germany\\
% ...
$^{28}$ Johannes Gutenberg University of Mainz, Johann-Joachim-Becher-Weg 45, D-55099 Mainz, Germany\\
% ...
$^{32}$ Liaoning Normal University, Dalian 116029, People's Republic of China\\
% ...
$^{49}$ State Key Laboratory of Particle Detection and Electronics, Beijing 100049, Hefei 230026, People's Republic of China\\
% ...
$^{57}$ University of Jinan, Jinan 250022, People's Republic of China\\
% ...
$^{63}$ University of Science and Technology of China, Hefei 230026, People's Republic of China\\
% ...
$^{66}$ University of Turin and INFN, INFN, I-10125, Turin, Italy\\
$^{67}$ Uppsala University, Box 516, SE-75120 Uppsala, Sweden\\
% ...
$^{c}$ Also at the Novosibirsk State University, Novosibirsk, 630090, Russia\\
% ...
}\end{center}


%% file: Paper.bbl
\begin{thebibliography}{32}%
\makeatletter
\providecommand \@ifxundefined [1]{%
 \@ifx{#1\undefined}
}%
\providecommand \@ifnum [1]{%
 \ifnum #1\expandafter \@firstoftwo
 \else \expandafter \@secondoftwo
 \fi
}%
\providecommand \@ifx [1]{%
 \ifx #1\expandafter \@firstoftwo
 \else \expandafter \@secondoftwo
 \fi
}%
\providecommand \natexlab [1]{#1}%
\providecommand \enquote  [1]{``#1''}%
\providecommand \bibnamefont  [1]{#1}%
\providecommand \bibfnamefont [1]{#1}%
\providecommand \citenamefont [1]{#1}%
\providecommand \href@noop [0]{\@secondoftwo}%
\providecommand \href [0]{\begingroup \@sanitize@url \@href}%
\providecommand \@href[1]{\@@startlink{#1}\@@href}%
\providecommand \@@href[1]{\endgroup#1\@@endlink}%
\providecommand \@sanitize@url [0]{\catcode `\\12\catcode `\$12\catcode
  `\&12\catcode `\#12\catcode `\^12\catcode `\_12\catcode `\%12\relax}%
\providecommand \@@startlink[1]{}%
\providecommand \@@endlink[0]{}%
\providecommand \url  [0]{\begingroup\@sanitize@url \@url }%
\providecommand \@url [1]{\endgroup\@href {#1}{\urlprefix }}%
\providecommand \urlprefix  [0]{URL }%
\providecommand \Eprint [0]{\href }%
\providecommand \doibase [0]{https://doi.org/}%
\providecommand \selectlanguage [0]{\@gobble}%
\providecommand \bibinfo  [0]{\@secondoftwo}%
\providecommand \bibfield  [0]{\@secondoftwo}%
\providecommand \translation [1]{[#1]}%
\providecommand \BibitemOpen [0]{}%
\providecommand \bibitemStop [0]{}%
\providecommand \bibitemNoStop [0]{.\EOS\space}%
\providecommand \EOS [0]{\spacefactor3000\relax}%
\providecommand \BibitemShut  [1]{\csname bibitem#1\endcsname}%
\let\auto@bib@innerbib\@empty
\bibitem [{\citenamefont {Renard}(1976)}]{Renard}%
  \BibitemOpen
  \bibfield  {author} {\bibinfo {author} {\bibfnamefont {F.}~\bibnamefont
  {Renard}},\ }\href
  {http://www.sciencedirect.com/science/article/pii/0370269376900204}
  {\bibfield  {journal} {\bibinfo  {journal} {Phys. Lett. B}\ }\textbf
  {\bibinfo {volume} {65}},\ \bibinfo {pages} {157 } (\bibinfo {year}
  {1976})}\BibitemShut {NoStop}%
\bibitem [{\citenamefont {Bodwin}\ \emph {et~al.}(1992)\citenamefont {Bodwin},
  \citenamefont {Braaten},\ and\ \citenamefont {Lepage}}]{Bodwin}%
  \BibitemOpen
  \bibfield  {author} {\bibinfo {author} {\bibfnamefont {G.~T.}\ \bibnamefont
  {Bodwin}}, \bibinfo {author} {\bibfnamefont {E.}~\bibnamefont {Braaten}},\
  and\ \bibinfo {author} {\bibfnamefont {G.~P.}\ \bibnamefont {Lepage}},\
  }\href {https://link.aps.org/doi/10.1103/PhysRevD.46.R1914} {\bibfield
  {journal} {\bibinfo  {journal} {Phys. Rev. D}\ }\textbf {\bibinfo {volume}
  {46}},\ \bibinfo {pages} {R1914} (\bibinfo {year} {1992})}\BibitemShut
  {NoStop}%
\bibitem [{\citenamefont {Novikov}\ \emph {et~al.}(1978)\citenamefont {Novikov}
  \emph {et~al.}}]{Novikov}%
  \BibitemOpen
  \bibfield  {author} {\bibinfo {author} {\bibfnamefont {V.}~\bibnamefont
  {Novikov}} \emph {et~al.},\ }\href
  {http://www.sciencedirect.com/science/article/pii/0370157378901205}
  {\bibfield  {journal} {\bibinfo  {journal} {Phys. Rep.}\ }\textbf {\bibinfo
  {volume} {41}},\ \bibinfo {pages} {1 } (\bibinfo {year} {1978})}\BibitemShut
  {NoStop}%
\bibitem [{\citenamefont {Kuang}\ \emph {et~al.}(1988)\citenamefont {Kuang},
  \citenamefont {Tuan},\ and\ \citenamefont {Yan}}]{Kuang}%
  \BibitemOpen
  \bibfield  {author} {\bibinfo {author} {\bibfnamefont {Y.-P.}\ \bibnamefont
  {Kuang}}, \bibinfo {author} {\bibfnamefont {S.~F.}\ \bibnamefont {Tuan}},\
  and\ \bibinfo {author} {\bibfnamefont {T.-M.}\ \bibnamefont {Yan}},\ }\href
  {https://link.aps.org/doi/10.1103/PhysRevD.37.1210} {\bibfield  {journal}
  {\bibinfo  {journal} {Phys. Rev. D}\ }\textbf {\bibinfo {volume} {37}},\
  \bibinfo {pages} {1210} (\bibinfo {year} {1988})}\BibitemShut {NoStop}%
\bibitem [{\citenamefont {Dissertori}\ \emph {et~al.}(2009)\citenamefont
  {Dissertori}, \citenamefont {Knowles},\ and\ \citenamefont
  {Schmelling}}]{dissertori2009quantum}%
  \BibitemOpen
  \bibfield  {author} {\bibinfo {author} {\bibfnamefont {G.}~\bibnamefont
  {Dissertori}}, \bibinfo {author} {\bibfnamefont {I.}~\bibnamefont
  {Knowles}},\ and\ \bibinfo {author} {\bibfnamefont {M.}~\bibnamefont
  {Schmelling}},\ }\href{https://global.oup.com/academic/product/quantum-chromodynamics-9780199566419?cc=de&lang=en&}
  {\bibfield  {journal} {\bibinfo  {journal} {ISBN: 9780199566419}\ }}
  {\bibinfo {series} {International Series of Monographs on Physics} (\bibinfo {year} {2009})}\BibitemShut {NoStop}%
\bibitem [{\citenamefont {Rubin}\ \emph {et~al.}(2005)\citenamefont {Rubin}
  \emph {et~al.}}]{PhysRevD.72.092004}%
  \BibitemOpen
  \bibfield  {author} {\bibinfo {author} {\bibfnamefont {P.}~\bibnamefont
  {Rubin}} \emph {et~al.} (\bibinfo {collaboration} {CLEO Collaboration}),\
  }\href {https://doi.org/10.1103/PhysRevD.72.092004} {\bibfield  {journal}
  {\bibinfo  {journal} {Phys. Rev. D}\ }\textbf {\bibinfo {volume} {72}},\
  \bibinfo {pages} {092004} (\bibinfo {year} {2005})}\BibitemShut {NoStop}%
\bibitem [{\citenamefont {Rosner}\ \emph {et~al.}(2005)\citenamefont {Rosner}
  \emph {et~al.}}]{PhysRevLett.95.102003}%
  \BibitemOpen
  \bibfield  {author} {\bibinfo {author} {\bibfnamefont {J.~L.}\ \bibnamefont
  {Rosner}} \emph {et~al.} (\bibinfo {collaboration} {CLEO Collaboration}),\
  }\href {https://doi.org/10.1103/PhysRevLett.95.102003} {\bibfield  {journal}
  {\bibinfo  {journal} {Phys. Rev. Lett.}\ }\textbf {\bibinfo {volume} {95}},\
  \bibinfo {pages} {102003} (\bibinfo {year} {2005})}\BibitemShut {NoStop}%
\bibitem [{\citenamefont {Andreotti}\ \emph {et~al.}(2005)\citenamefont
  {Andreotti} \emph {et~al.}}]{PhysRevD.72.032001}%
  \BibitemOpen
  \bibfield  {author} {\bibinfo {author} {\bibfnamefont {M.}~\bibnamefont
  {Andreotti}} \emph {et~al.} (\bibinfo {collaboration} {Fermilab E835
  Collaboration}),\ }\href {https://doi.org/10.1103/PhysRevD.72.032001}
  {\bibfield  {journal} {\bibinfo  {journal} {Phys. Rev. D}\ }\textbf {\bibinfo
  {volume} {72}},\ \bibinfo {pages} {032001} (\bibinfo {year}
  {2005})}\BibitemShut {NoStop}%
\bibitem [{\citenamefont {Dobbs}\ \emph {et~al.}(2008)\citenamefont {Dobbs}
  \emph {et~al.}}]{PhysRevLett.101.182003}%
  \BibitemOpen
  \bibfield  {author} {\bibinfo {author} {\bibfnamefont {S.}~\bibnamefont
  {Dobbs}} \emph {et~al.} (\bibinfo {collaboration} {CLEO Collaboration}),\
  }\href {https://doi.org/10.1103/PhysRevLett.101.182003} {\bibfield  {journal}
  {\bibinfo  {journal} {Phys. Rev. Lett.}\ }\textbf {\bibinfo {volume} {101}},\
  \bibinfo {pages} {182003} (\bibinfo {year} {2008})}\BibitemShut {NoStop}%
\bibitem [{\citenamefont {Ablikim}\ \emph
  {et~al.}(2010{\natexlab{a}})\citenamefont {Ablikim} \emph
  {et~al.}}]{PhysRevLett.104.132002}%
  \BibitemOpen
  \bibfield  {author} {\bibinfo {author} {\bibfnamefont {M.}~\bibnamefont
  {Ablikim}} \emph {et~al.} (\bibinfo {collaboration} {BESIII Collaboration}),\
  }\href {https://doi.org/10.1103/PhysRevLett.104.132002} {\bibfield  {journal}
  {\bibinfo  {journal} {Phys. Rev. Lett.}\ }\textbf {\bibinfo {volume} {104}},\
  \bibinfo {pages} {132002} (\bibinfo {year} {2010}{\natexlab{a}})}\BibitemShut
  {NoStop}%
\bibitem [{\citenamefont {Zyla}\ \emph {et~al.}(2020)\citenamefont {Zyla} \emph
  {et~al.}}]{pdg}%
  \BibitemOpen
  \bibfield  {author} {\bibinfo {author} {\bibfnamefont {P.~A.}\ \bibnamefont
  {Zyla}} \emph {et~al.} (\bibinfo {collaboration} {Particle Data Group}),\
  }\bibfield  {title} {\bibinfo {title} {{Review of Particle Physics}},\
  }\bibfield  {journal} {\bibinfo  {journal} {Progress of Theoretical and
  Experimental Physics}\ }\textbf {\bibinfo {volume} {2020}},\ \href
  {https://doi.org/10.1093/ptep/ptaa104} {10.1093/ptep/ptaa104} (\bibinfo
  {year} {2020}),\ \bibinfo {note} {083C01}\BibitemShut {NoStop}%
\bibitem [{\citenamefont {Ablikim}\ \emph
  {et~al.}(2019{\natexlab{a}})\citenamefont {Ablikim} \emph
  {et~al.}}]{hctohadrons}%
  \BibitemOpen
  \bibfield  {author} {\bibinfo {author} {\bibfnamefont {M.}~\bibnamefont
  {Ablikim}} \emph {et~al.} (\bibinfo {collaboration} {BESIII Collaboration}),\
  }\href {https://doi.org/10.1103/PhysRevD.99.072008} {\bibfield  {journal}
  {\bibinfo  {journal} {Phys. Rev. D}\ }\textbf {\bibinfo {volume} {99}},\
  \bibinfo {pages} {072008} (\bibinfo {year} {2019}{\natexlab{a}})}\BibitemShut
  {NoStop}%
\bibitem [{\citenamefont {Adams}\ \emph {et~al.}(2009)\citenamefont {Adams}
  \emph {et~al.}}]{PhysRevD.80.051106}%
  \BibitemOpen
  \bibfield  {author} {\bibinfo {author} {\bibfnamefont {G.~S.}\ \bibnamefont
  {Adams}} \emph {et~al.} (\bibinfo {collaboration} {CLEO Collaboration}),\
  }\href {https://doi.org/10.1103/PhysRevD.80.051106} {\bibfield  {journal}
  {\bibinfo  {journal} {Phys. Rev. D}\ }\textbf {\bibinfo {volume} {80}},\
  \bibinfo {pages} {051106} (\bibinfo {year} {2009})}\BibitemShut {NoStop}%
\bibitem [{\citenamefont {Ablikim}\ \emph {et~al.}(2018)\citenamefont {Ablikim}
  \emph {et~al.}}]{psip_events}%
  \BibitemOpen
  \bibfield  {author} {\bibinfo {author} {\bibfnamefont {M.}~\bibnamefont
  {Ablikim}} \emph {et~al.} (\bibinfo {collaboration} {BESIII Collaboration}),\
  }\href {https://doi.org/10.1088/1674-1137/42/2/023001} {\bibfield  {journal}
  {\bibinfo  {journal} {Chin. Phys. C}\ }\textbf {\bibinfo {volume} {42}},\
  \bibinfo {pages} {023001} (\bibinfo {year} {2018})}\BibitemShut {NoStop}%
\bibitem [{\citenamefont {Ablikim}\ \emph
  {et~al.}(2010{\natexlab{b}})\citenamefont {Ablikim} \emph
  {et~al.}}]{BES_Detector}%
  \BibitemOpen
  \bibfield  {author} {\bibinfo {author} {\bibfnamefont {M.}~\bibnamefont
  {Ablikim}} \emph {et~al.} (\bibinfo {collaboration} {BESIII Collaboration}),\
  }\href {https://doi.org/10.1016/j.nima.2009.12.050} {\bibfield  {journal}
  {\bibinfo  {journal} {Nucl. Instrum. Methods Phys. A}\ }\textbf {\bibinfo
  {volume} {614}},\ \bibinfo {pages} {345} (\bibinfo {year}
  {2010}{\natexlab{b}})}\BibitemShut {NoStop}%
\bibitem [{\citenamefont {Yu}\ \emph {et~al.}(2016)\citenamefont {Yu} \emph
  {et~al.}}]{Yu:IPAC2016-TUYA01}%
  \BibitemOpen
  \bibfield  {author} {\bibinfo {author} {\bibfnamefont {C.}~\bibnamefont {Yu}}
  \emph {et~al.},\ } {\bibfield  {journal} {\bibinfo  {journal} {in Proceedings of IPAC2016,
  Busan, Korea}}}\ ,\ \href {http://jacow.org/ipac2016/papers/tuya01.pdf}{\bibinfo {pages} {1014} (\bibinfo {year} {2016})}\BibitemShut {NoStop}%
\bibitem [{\citenamefont {Lange}(2001)}]{evtgen1}%
  \BibitemOpen
  \bibfield  {author} {\bibinfo {author} {\bibfnamefont {D.~J.}\ \bibnamefont
  {Lange}},\ }\href {https://doi.org/10.1016/S0168-9002(01)00089-4} {\bibfield
  {journal} {\bibinfo  {journal} {Nucl. Instrum. Methods Phys. A}\ }\textbf
  {\bibinfo {volume} {462}},\ \bibinfo {pages} {152} (\bibinfo {year}
  {2001})}\BibitemShut {NoStop}%
\bibitem [{\citenamefont {Ping}(2008)}]{evtgen2}%
  \BibitemOpen
  \bibfield  {author} {\bibinfo {author} {\bibfnamefont {R.~G.}\ \bibnamefont
  {Ping}},\ }\href {https://doi.org/10.1088/1674-1137/32/8/001} {\bibfield
  {journal} {\bibinfo  {journal} {Chin. Phys. C}\ }\textbf {\bibinfo {volume}
  {32}},\ \bibinfo {pages} {599} (\bibinfo {year} {2008})}\BibitemShut
  {NoStop}%
\bibitem [{\citenamefont {Jadach}\ \emph {et~al.}(2001)\citenamefont {Jadach},
  \citenamefont {Ward},\ and\ \citenamefont {Wa\ifmmode~\mbox{\c{}}\else
  \c{}\fi{}s}}]{kkmc1}%
  \BibitemOpen
  \bibfield  {author} {\bibinfo {author} {\bibfnamefont {S.}~\bibnamefont
  {Jadach}}, \bibinfo {author} {\bibfnamefont {B.~F.~L.}\ \bibnamefont
  {Ward}},\ and\ \bibinfo {author} {\bibfnamefont {Z.}~\bibnamefont
  {Wa\ifmmode~\mbox{\c{}}\else \c{}\fi{}s}},\ }\href
  {https://doi.org/10.1103/PhysRevD.63.113009} {\bibfield  {journal} {\bibinfo
  {journal} {Phys. Rev. D}\ }\textbf {\bibinfo {volume} {63}},\ \bibinfo
  {pages} {113009} (\bibinfo {year} {2001})}\BibitemShut {NoStop}%
\bibitem [{\citenamefont {Jadach}\ \emph {et~al.}(2000)\citenamefont {Jadach},
  \citenamefont {Ward},\ and\ \citenamefont {Wa\ifmmode~\mbox{\c{}}\else
  \c{}\fi{}s}}]{kkmc2}%
  \BibitemOpen
  \bibfield  {author} {\bibinfo {author} {\bibfnamefont {S.}~\bibnamefont
  {Jadach}}, \bibinfo {author} {\bibfnamefont {B.~F.~L.}\ \bibnamefont
  {Ward}},\ and\ \bibinfo {author} {\bibfnamefont {Z.}~\bibnamefont
  {Wa\ifmmode~\mbox{\c{}}\else \c{}\fi{}s}},\ }\href
  {https://doi.org/https://doi.org/10.1016/S0010-4655(00)00048-5} {\bibfield
  {journal} {\bibinfo  {journal} {Comput. Phys. Commun.}\ }\textbf {\bibinfo
  {volume} {130}},\ \bibinfo {pages} {260 } (\bibinfo {year}
  {2000})}\BibitemShut {NoStop}%
\bibitem [{\citenamefont {Patrignani}\ \emph {et~al.}(2016)\citenamefont
  {Patrignani} \emph {et~al.}}]{pdg16}%
  \BibitemOpen
  \bibfield  {author} {\bibinfo {author} {\bibfnamefont {C.}~\bibnamefont
  {Patrignani}} \emph {et~al.} (\bibinfo {collaboration} {Particle Data
  Group}),\ }\bibfield  {title} {\bibinfo {title} {{Review of Particle
  Physics}},\ }\href {https://doi.org/10.1088/1674-1137/40/10/100001}
  {\bibfield  {journal} {\bibinfo  {journal} {Chin. Phys. C}\ }\textbf
  {\bibinfo {volume} {40}},\ \bibinfo {pages} {100001} (\bibinfo {year}
  {2016})}\BibitemShut {NoStop}%
\bibitem [{\citenamefont {Chen}\ \emph {et~al.}(2000)\citenamefont {Chen},
  \citenamefont {Huang}, \citenamefont {Qi}, \citenamefont {Zhang},\ and\
  \citenamefont {Zhu}}]{lundcharm}%
  \BibitemOpen
  \bibfield  {author} {\bibinfo {author} {\bibfnamefont {J.~C.}\ \bibnamefont
  {Chen}}, \bibinfo {author} {\bibfnamefont {G.~S.}\ \bibnamefont {Huang}},
  \bibinfo {author} {\bibfnamefont {X.~R.}\ \bibnamefont {Qi}}, \bibinfo
  {author} {\bibfnamefont {D.~H.}\ \bibnamefont {Zhang}},\ and\ \bibinfo
  {author} {\bibfnamefont {Y.~S.}\ \bibnamefont {Zhu}},\ }\href
  {https://doi.org/10.1103/PhysRevD.62.034003} {\bibfield  {journal} {\bibinfo
  {journal} {Phys. Rev. D}\ }\textbf {\bibinfo {volume} {62}},\ \bibinfo
  {pages} {034003} (\bibinfo {year} {2000})}\BibitemShut {NoStop}%
\bibitem [{\citenamefont {Das}()}]{Das2016}%
  \BibitemOpen
  \bibfield  {author} {\bibinfo {author} {\bibfnamefont {S.}~\bibnamefont
  {Das}},\ }\href@noop {} {\ }\Eprint {https://arxiv.org/abs/1603.08591}
  {arXiv:1603.08591 [hep-ex]} \BibitemShut {NoStop}%
\bibitem [{\citenamefont {Albrecht}\ \emph {et~al.}(1994)\citenamefont
  {Albrecht} \emph {et~al.}}]{ALBRECHT1994217}%
  \BibitemOpen
  \bibfield  {author} {\bibinfo {author} {\bibfnamefont {H.}~\bibnamefont
  {Albrecht}} \emph {et~al.} (\bibinfo {collaboration} {ARGUS Collaboration}),\
  }\href {https://doi.org/https://doi.org/10.1016/0370-2693(94)01302-0}
  {\bibfield  {journal} {\bibinfo  {journal} {Phys. Lett. B}\ }\textbf
  {\bibinfo {volume} {340}},\ \bibinfo {pages} {217 } (\bibinfo {year}
  {{1994}})}\BibitemShut {NoStop}%
\bibitem [{\citenamefont {Demortier}(2002)}]{Demortier:2002ic}%
  \BibitemOpen
  \bibfield  {author} {\bibinfo {author} {\bibfnamefont {L.}~\bibnamefont
  {Demortier}},\ } {\bibfield  {journal} {\bibinfo  {journal} {Proceedings of the Advanced Statistical Techniques in Particle Physics}\ }}
  \href{https://www.ippp.dur.ac.uk/Workshops/02/statistics/proceedings/demortier.pdf}
  {\textbf {\bibinfo {volume}
  {0203181}},\ \bibinfo {pages} {145} (\bibinfo {year} {2002})}\BibitemShut
  {NoStop}%
\bibitem [{\citenamefont {Yuan}\ \emph {et~al.}(2016)\citenamefont {Yuan} \emph
  {et~al.}}]{p_tracking}%
  \BibitemOpen
  \bibfield  {author} {\bibinfo {author} {\bibfnamefont {W.~L.}\ \bibnamefont
  {Yuan}} \emph {et~al.},\ }\href
  {https://doi.org/10.1088/1674-1137/40/2/026201} {\bibfield  {journal}
  {\bibinfo  {journal} {Chin. Phys. C}\ }\textbf {\bibinfo {volume} {40}},\
  \bibinfo {pages} {026201} (\bibinfo {year} {2016})}\BibitemShut {NoStop}%
\bibitem [{\citenamefont {Ablikim}\ \emph
  {et~al.}(2019{\natexlab{b}})\citenamefont {Ablikim} \emph
  {et~al.}}]{k_pi_tracking}%
  \BibitemOpen
  \bibfield  {author} {\bibinfo {author} {\bibfnamefont {M.}~\bibnamefont
  {Ablikim}} \emph {et~al.} (\bibinfo {collaboration} {BESIII Collaboration}),\
  }\href {https://doi.org/10.1103/PhysRevD.99.091103} {\bibfield  {journal}
  {\bibinfo  {journal} {Phys. Rev. D}\ }\textbf {\bibinfo {volume} {99}},\
  \bibinfo {pages} {091103} (\bibinfo {year} {2019}{\natexlab{b}})}\BibitemShut
  {NoStop}%
\bibitem [{\citenamefont {Ablikim}\ \emph {et~al.}(2012)\citenamefont {Ablikim}
  \emph {et~al.}}]{pion_PID}%
  \BibitemOpen
  \bibfield  {author} {\bibinfo {author} {\bibfnamefont {M.}~\bibnamefont
  {Ablikim}} \emph {et~al.} (\bibinfo {collaboration} {BESIII Collaboration}),\
  }\href {https://doi.org/10.1103/PhysRevD.86.092009} {\bibfield  {journal}
  {\bibinfo  {journal} {Phys. Rev. D}\ }\textbf {\bibinfo {volume} {86}},\
  \bibinfo {pages} {092009} (\bibinfo {year} {2012})}\BibitemShut {NoStop}%
\bibitem [{\citenamefont {Ablikim}\ \emph
  {et~al.}(2010{\natexlab{c}})\citenamefont {Ablikim} \emph
  {et~al.}}]{photon_effi}%
  \BibitemOpen
  \bibfield  {author} {\bibinfo {author} {\bibfnamefont {M.}~\bibnamefont
  {Ablikim}} \emph {et~al.} (\bibinfo {collaboration} {BESIII Collaboration}),\
  }\href {https://doi.org/10.1103/PhysRevD.81.052005} {\bibfield  {journal}
  {\bibinfo  {journal} {Phys. Rev. D}\ }\textbf {\bibinfo {volume} {81}},\
  \bibinfo {pages} {052005} (\bibinfo {year} {2010}{\natexlab{c}})}\BibitemShut
  {NoStop}%
\bibitem [{\citenamefont {Ablikim}\ \emph
  {et~al.}(2019{\natexlab{c}})\citenamefont {Ablikim} \emph
  {et~al.}}]{ks_reco}%
  \BibitemOpen
  \bibfield  {author} {\bibinfo {author} {\bibfnamefont {M.}~\bibnamefont
  {Ablikim}} \emph {et~al.} (\bibinfo {collaboration} {BESIII Collaboration}),\
  }\href {https://doi.org/10.1103/PhysRevD.99.032002} {\bibfield  {journal}
  {\bibinfo  {journal} {Phys. Rev. D}\ }\textbf {\bibinfo {volume} {99}},\
  \bibinfo {pages} {032002} (\bibinfo {year} {2019}{\natexlab{c}})}\BibitemShut
  {NoStop}%
\bibitem [{\citenamefont {Ablikim}\ \emph {et~al.}(2013)\citenamefont {Ablikim}
  \emph {et~al.}}]{helix_cor}%
  \BibitemOpen
  \bibfield  {author} {\bibinfo {author} {\bibfnamefont {M.}~\bibnamefont
  {Ablikim}} \emph {et~al.} (\bibinfo {collaboration} {BESIII Collaboration}),\
  }\href {https://doi.org/10.1103/PhysRevD.87.012002} {\bibfield  {journal}
  {\bibinfo  {journal} {Phys. Rev. D}\ }\textbf {\bibinfo {volume} {87}},\
  \bibinfo {pages} {012002} (\bibinfo {year} {2013})}\BibitemShut {NoStop}%
\bibitem [{\citenamefont {Ablikim}\ \emph {et~al.}(2020)\citenamefont {Ablikim}
  \emph {et~al.}}]{Ablikim:2019hff}%
  \BibitemOpen
  \bibfield  {author} {\bibinfo {author} {\bibfnamefont {M.}~\bibnamefont
  {Ablikim}} \emph {et~al.} (\bibinfo {collaboration} {BESIII Collaboration}),\
  }\href {https://doi.org/10.1088/1674-1137/44/4/040001} {\bibfield  {journal}
  {\bibinfo  {journal} {Chin. Phys. C}\ }\textbf {\bibinfo {volume} {44}},\
  \bibinfo {pages} {040001} (\bibinfo {year} {2020})}\BibitemShut {NoStop}%
\end{thebibliography}%
